\documentclass[fleqn,usenatbib]{mnras}

\usepackage{newtxtext,newtxmath}
\usepackage[T1]{fontenc}

\DeclareRobustCommand{\VAN}[3]{#2}
\let\VANthebibliography\thebibliography
\def\thebibliography{\DeclareRobustCommand{\VAN}[3]{##3}\VANthebibliography}

\usepackage{graphicx}	% Including figure files
\usepackage{stfloats}
\usepackage{float}
\usepackage{lipsum}
\usepackage[normalem]{ulem}
\usepackage{subfig}

\newcommand{\RNum}[1]{\uppercase\expandafter{\romannumeral #1\relax}}
\newcommand{\msun}{M\textsubscript{\(\odot\)}}

\title[Non-Keplerian spirals in HD142527]{Non-Keplerian spirals, a gas-pressure dust trap and an eccentric gas cavity in the circumbinary disc around HD 142527}

\author[H. Garg et al.]{
H. Garg$^{1}$\thanks{E-mail: himanshi.garg@monash.edu}, C. Pinte$^{1,2}$, V. Christiaens$^{1}$, D. J. Price$^{1}$, J. S. Lazendic$^{1}$, Y. Boehler$^{2}$, \newauthor S. Casassus$^{3}$, S. Marino$^{4}$, S. Perez$^{5}$ and A. Zuleta$^{3}$
\\
$^{1}$School of Physics and Astronomy, Monash University, Clayton VIC 3800, Australia\\
$^{2}$Univ. Grenoble Alpes, CNRS, IPAG, F-38000 Grenoble, France\\
$^{3}$Departamento de Astronomía, Universidad de Chile, Casilla 36-D, Santiago, Chile\\
$^{4}$Institute of Astronomy, University of Cambridge, Madingley Road, Cambridge CB3 0HA, UK\\
$^{5}$Universidad de Santiago de Chile, Av. Libertador Bernardo O'Higgins 3363, Estación Central, Santiago, Chile
}

\date{Accepted XXX. Received YYY; in original form ZZZ}

\pubyear{2021}

\begin{document}
\label{firstpage}
\pagerange{\pageref{firstpage}--\pageref{lastpage}}
\maketitle

\begin{abstract}
We present ALMA observations of the $^{12}$CO, $^{13}$CO, C$^{18}$O J=2-1 transitions and the 1.3\,mm continuum emission for the circumbinary disc around HD 142527, at an angular resolution of $\approx$\,0\farcs3. We observe multiple spiral structures in intensity, velocity and velocity dispersion for the $^{12}$CO and $^{13}$CO gas tracers. A newly detected $^{12}$CO spiral originates from the dust horseshoe, and is rotating at super-Keplerian velocity or vertically ascending, whilst the inter-spiral gas is rotating at sub-Keplerian velocities. This new spiral possibly connects to a previously identified spiral, thus spanning > 360$^\circ$. A spatial offset of ~30 au is observed between the $^{12}$CO and $^{13}$CO spirals, to which we hypothesize that the gas layers are propagating at different speeds (``surfing'') due to a non-zero vertical temperature gradient. Leveraging the varying optical depths between the CO isotopologues, we reconstruct temperature and column density maps of the outer disc. Gas surface density peaks at r\,$\approx$\,180\,au, coincident with the peak of continuum emission. Here the dust grains have a Stokes number of $\approx$\,1, confirming radial and azimuthal trapping in the horseshoe. We measure a cavity radius at half-maximum surface density of $\approx$\,100\,au, and a cavity eccentricity between 0.3 and 0.45.

\end{abstract}

\begin{keywords}
accretion discs -- circumstellar matter -- submillimetre: planetary systems -- stars: individual: HD142527
\end{keywords}

%%%%%%%%%%%%%%%%%%%%%%%%%%%%%%%%%%%%%%%%%%%%%%%%%%

%%%%%%%%%%%%%%%%% BODY OF PAPER %%%%%%%%%%%%%%%%%%

\section{Introduction}
\label{sec:intro}

\begin{figure*}
    %\centering
    \includegraphics[width=\textwidth]{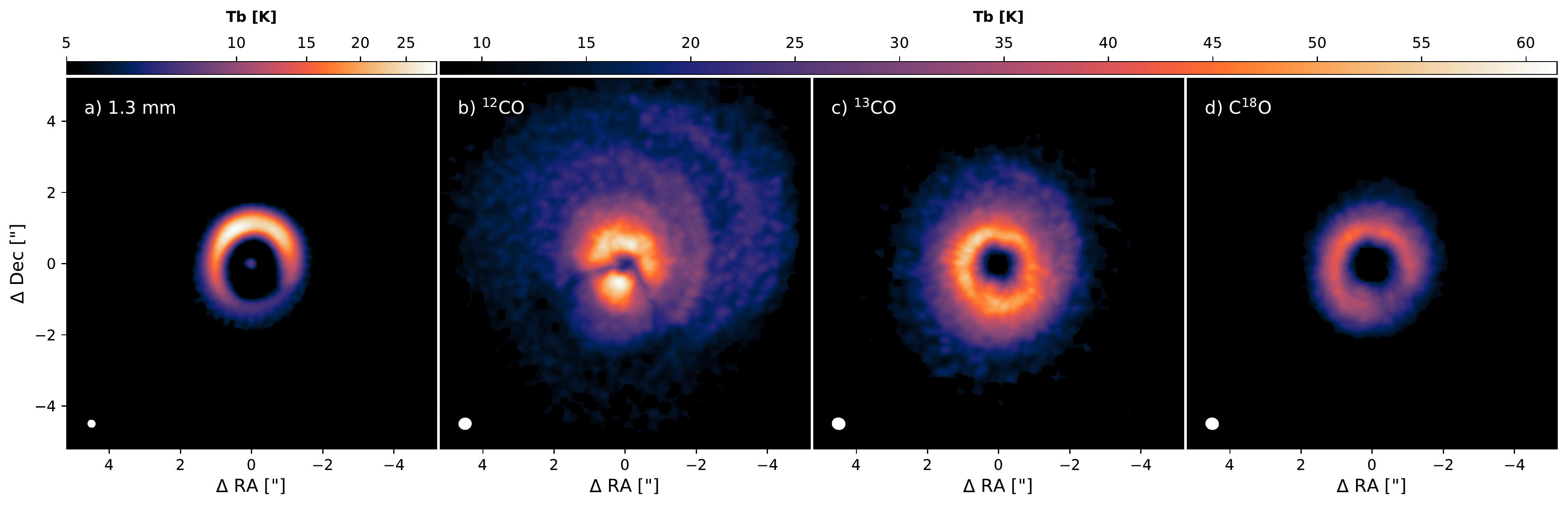}
    \caption{\textbf{Spirals in $^{12}$CO and $^{13}$CO.} 1.3mm continuum and non-continuum subtracted peak brightness temperature maps for $^{12}$CO, $^{13}$CO and C$^{18}$O. The disc is inclined about a position angle of -20$^{\circ}$ such that the NE side is the far side and SW is the near side. The sharp drop in $^{12}$CO emission along an inverted V-shaped feature (e.g. at PAs $\approx$100$^{\circ}$ and 200$^{\circ}$) is due to the intervening foreground cloud \citep{Casassus2013b}. The continuum map is shown with a square root color stretch, while the CO maps are shown with the same linear color scale. Respective beams are shown on the bottom left of each panel.}
    \label{fig:original}
\end{figure*}

Protoplanetary discs with large central cavities are believed to be the result of interactions with internal companion(s). The circumbinary disc around HD 142527, a Herbig Fe pre-main-sequence (PMS) star, is one such example. The outer disc is oriented nearly face-on, with estimated inclinations ranging from 20 to 28$^{\circ}$ about a position angle of -20$^{\circ}$ \citep{Verhoeff2011,Avenhaus2014,Perez2015}, and located at a distance of 157.3 $\pm$ 1.2 pc \citep{Gaia2018}. The system hosts a large (r$\approx$140 au) central dust cavity \citep{Avenhaus2017, Casassus2013}; a distinct dust horseshoe \citep{Ohashi2008, Casassus2013}; large scale spiral arm structures seen in both near-IR \citep{Fukagawa2006, Casassus2012} and millimetre-wavelength observations \citep{Christiaens2014}; and an inclined inner disc casting shadows on the outer disc \citep{Marino2015}, revealed by non-Keplerian kinematics inside the cavity \citep{Casassus2015,Rosenfeld2014}.

\cite{Fukagawa2006} first speculated the presence of a companion as the explanation for the observed large scale spiral arm structures. \cite{Biller2012} provided the first direct evidence for a possible low-mass companion in the inner part of the cavity. Later works refined the companion properties and orbital parameters \citep{Close2014, Lacour2016,Christiaens2018}, in particular the semi-major axis ranging from 12 to 31\,au \citep{Claudi2019}.

Millimetre-emitting dust grains are seen to be depleted inside the cavity whilst up to centimetre sized grains are detected within the radial and azimuthal dust trap surrounding the cavity \citep{Casassus2015b}. \cite{Perez2015} reported a shallower gas depletion inside the central cavity with $^{12}$CO found to be optically thick throughout. \cite{Price2018} demonstrated that the interplay between the binary and enclosing outer disc can explain most of the observed peculiar features in this disc: dust asymmetry as a result of an eccentric binary; spirals along the edge of the dust cavity; a central cavity due to dynamical clearing; and shadows.

In an effort to characterise the large scale physical and kinematic structure of the outer disc, we present deep high spectral ($\approx 85$\,m/s) and intermediate spatial ($\approx$ 0\farcs3) resolution ALMA observations of the 1.3\,mm continuum emission and the $^{12}$CO, $^{13}$CO and C$^{18}$O J=2-1 lines.

\section{Observation \& Calibration}

HD 142527 was observed using the Atacama Large Millimetre/sub-millimetre Array (ALMA) as part of Project 2015.1.01353.S. Executions were made on March 2 2016 (using 38 antennas) and July 26 2016 (using 40 antennas) in Band 6, with on source integration times of $\approx$ 35 and 58 minutes respectively. The continuum spectral window (spw) centred at 231.9GHz with a total bandwidth of 2GHz. $^{12}$CO, $^{13}$CO and C$^{18}$O spectral windows centred at 230.503GHz, 220.479GHz and 219.641GHz respectively; all covering a total bandwidth of 117.187 MHz with 1920 channels of 61.035 kHz channel width each ($\sim$84 m/s). J1427-4206 was used as the flux and bandpass calibrator, J1604-4441 as the phase calibrator, and J1604-4228 as the gain calibrator.

We applied the standard ALMA pipeline calibration using {\sc casa} 4.5.3 to both executions. At first, only the short-baseline data sets were self-calibrated. Prior to merging the self-calibrated short-baseline data with the non-self-calibrated long-baseline data (using {\sf concat}) both data sets were phase centered to a common pointing direction. We then self-calibrated the combined data sets. We performed a series of phase-only self-calibrations on continuum for solutions intervals of [$\sim$7 mins, 60, 30, 15 and 6 s]; stepping down sequentially and where the first solution interval is equal to the length of an entire scan. Following each iteration, we imaged the short baseline data to ensure peak SNR increased by > 5\% from the previous iteration. An additional amplitude+phase self-calibration was applied for a solution interval of $\sim$7 mins. The peak SNR increased by a factor of 11. The outlined self-calibration process was repeated on the continuum combined data set, where peak SNR increased by a factor of 9. We applied the generated phase and amplitude calibration tables to each line emission. The resulting noise for $^{12}$CO, $^{13}$CO and C$^{18}$O is 5.9 mJy beam$^{-1}$ for a 0\farcs33 $\times$ 0\farcs30 beam; 6.3 mJy beam$^{-1}$ for a 0\farcs35 $\times$ 0\farcs31 beam; and 4.1 mJy beam$^{-1}$ for a 0\farcs34 $\times$ 0\farcs31 beam, respectively. Peak SNR: $^{12}$CO\,$\approx$\,40, $^{13}$CO\,$\approx$\,33 and C$^{18}$O\,$\approx$\,39. We imaged the spectral cubes with velocity channel widths of 84\,m/s, i.e. at the native spectral resolution of the data.

All imaging was performed using the {\sc casa} task {\sf tclean} with the {\sf multi-scale} deconvolver and {\sf Briggs} weighting, robust=0.5. We used {\sf auto-multithresh} masking for both continuum and line cube data with a minimal buffer ({\sf smoothFactor}=0.05 and {\sf cutThreshold}=0.01) to avoid capturing noise in the model. The resulting continuum noise is 51\,$\mu$Jy beam$^{-1}$ for a beam size of 0\farcs30 $\times$ 0\farcs27; corresponding to a peak SNR\,$\approx$\,1300. Additionally, we also imaged the continuum using {\sf super-uniform} weighting, for placing constraints on the inner disc size, with a resultant noise of 66 $\mu$Jy beam$^{-1}$ for a beam size of 0\farcs19 $\times$ 0\farcs18 and peak SNR\,$\approx$\,500 (Fig.~\ref{fig:original}). Furthermore, we imaged the $^{12}$CO emission with uniform weighting for analysing the twisted gas kinematics within the cavity. The uniform weighted $^{12}$CO cube has a beam size of 0\farcs23 $\times$ 0\farcs25, with a noise level of 10 mJy beam$^{-1}$ and peak SNR\,$\approx$\,15. All moment maps are integrated using the {\sc casa} task {\sf immoments} over a velocity range of -1.00 to 8.24 km/s. Non-continuum subtracted integrated flux (moment 0) maps for $^{12}$CO, $^{13}$CO and C$^{18}$O have a noise level of 1.74, 2.19 and 1.28 mJy beam$^{-1}$ with peak SNRs of 460, 332 and 483, respectively.

\section{Results and Discussion}

\subsection{Twisted kinematics near the inner disc}

The inner disc remains unresolved in our super-uniform weighted continuum map (Fig.~\ref{fig:original}a), which has a resolution of 0\farcs19 $\times$ 0\farcs18. At this resolution, we can place an upper limit of $\sim$30\,au for the diameter of the dust component of the inner disc, along with a total flux of $2.0\pm0.2$\,mJy. The disc centre was located by fitting a Gaussian to this unresolved inner disc's continuum emission.

\begin{figure*}
    \centering
    \includegraphics[width=\textwidth]{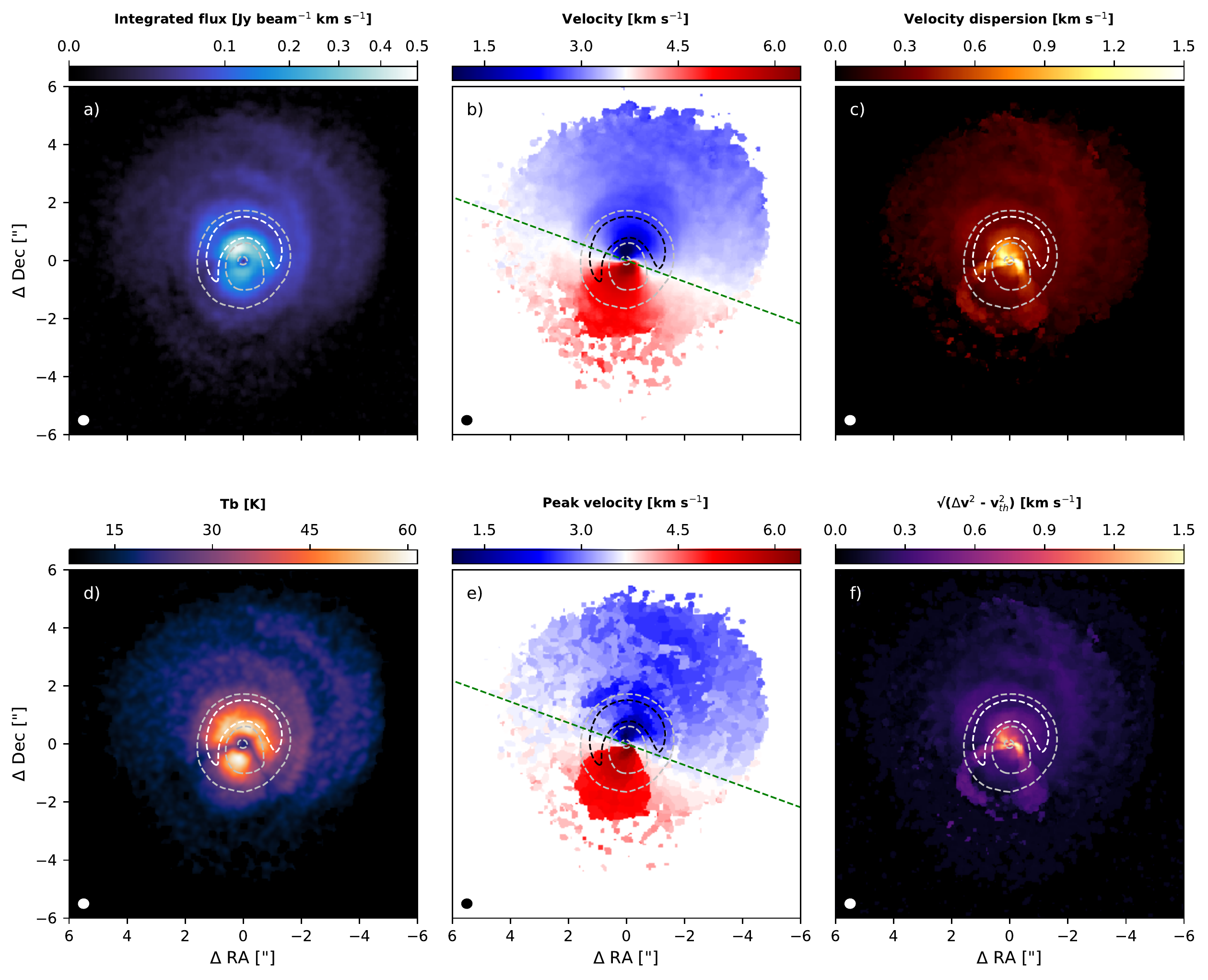}
    \caption{\textbf{Moment maps.} \textbf{a)}--\textbf{f}) moment 0, moment 1, moment 2, peak brightness temperature, peak velocity and velocity dispersion minus thermal component for $^{12}$CO, respectively. All integrated maps, with the exception of peak brightness temperature, are continuum subtracted. The continuum contour at 13.6 mJy beam$^{-1}$ (0.2 $\times$ maximum) is given in \textit{white} (in panels \textbf{a}, \textbf{c}, \textbf{d}, \textbf{f}) and \textit{black} (in panels \textbf{b} and \textbf{e}). A secondary continuum contour at 1 mJy beam$^{-1}$ ($\sim$20$\sigma$) is given in \textit{silver}. The \textit{green} dashed lines in panels \textbf{b} and \textbf{e} represent the rotation axis of the outer disc. Beam is given on the bottom left corner. The features seen towards the SE and SW directions (PAs $\approx$110$^{\circ}$ and 190$^{\circ}$) in panels \textbf{c}, \textbf{d}, \textbf{e} and \textbf{f} are due to the intervening cloud \citep{Casassus2013b}; an attempt at cloud correction is shown in Appendix~\ref{sec:appendix} along with a description on the procedure.}
    \label{fig:momentmaps}
\end{figure*}

\begin{figure}
    %\centering
    \includegraphics[width=\columnwidth]{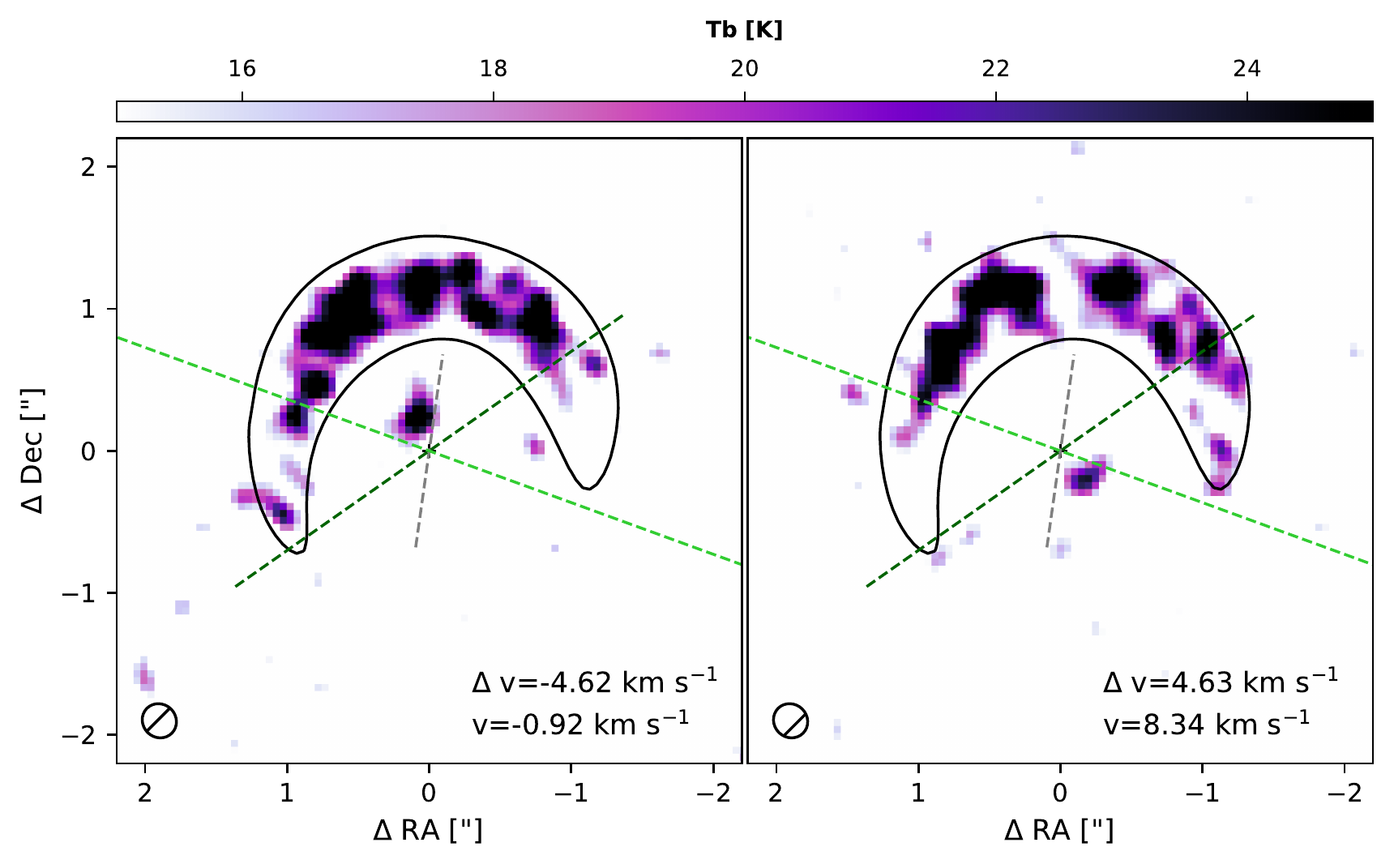}
    \caption{\textbf{Central gas twist and misalignment of the inner and outer discs.} High velocity channels (left:blue-shifted, right:red-shifted) of $^{12}$CO imaged using uniform weighting with emission above T$_\mathrm{b}$\,=\,15K, and without continuum subtraction. The systemic velocity is 3.70 km.s$^{-1}$. The cross marks the peak of the continuum for the inner disc. The \textit{black} solid line contour represents the 13.6 mJy beam$^{-1}$ (0.2 $\times$ maximum) continuum emission level. The \textit{dark green} dashed line represents the rotation axis of the central gas cavity at these high velocities \emph{i.e.} the axis perpendicular to the imaginary line connecting the two 'blobs' of gas emission at these high velocities.; the \textit{light green} line represent the average rotation axis of the outer disc; and the \textit{grey} dashed line represents the position angle of the inner disc as deduced from shadows in scattered light \citep{Marino2015}. Beam is given on the bottom left.}
    \label{fig:twist}
\end{figure}

\begin{figure*}
    %\centering
    \includegraphics[width=0.87\textwidth]{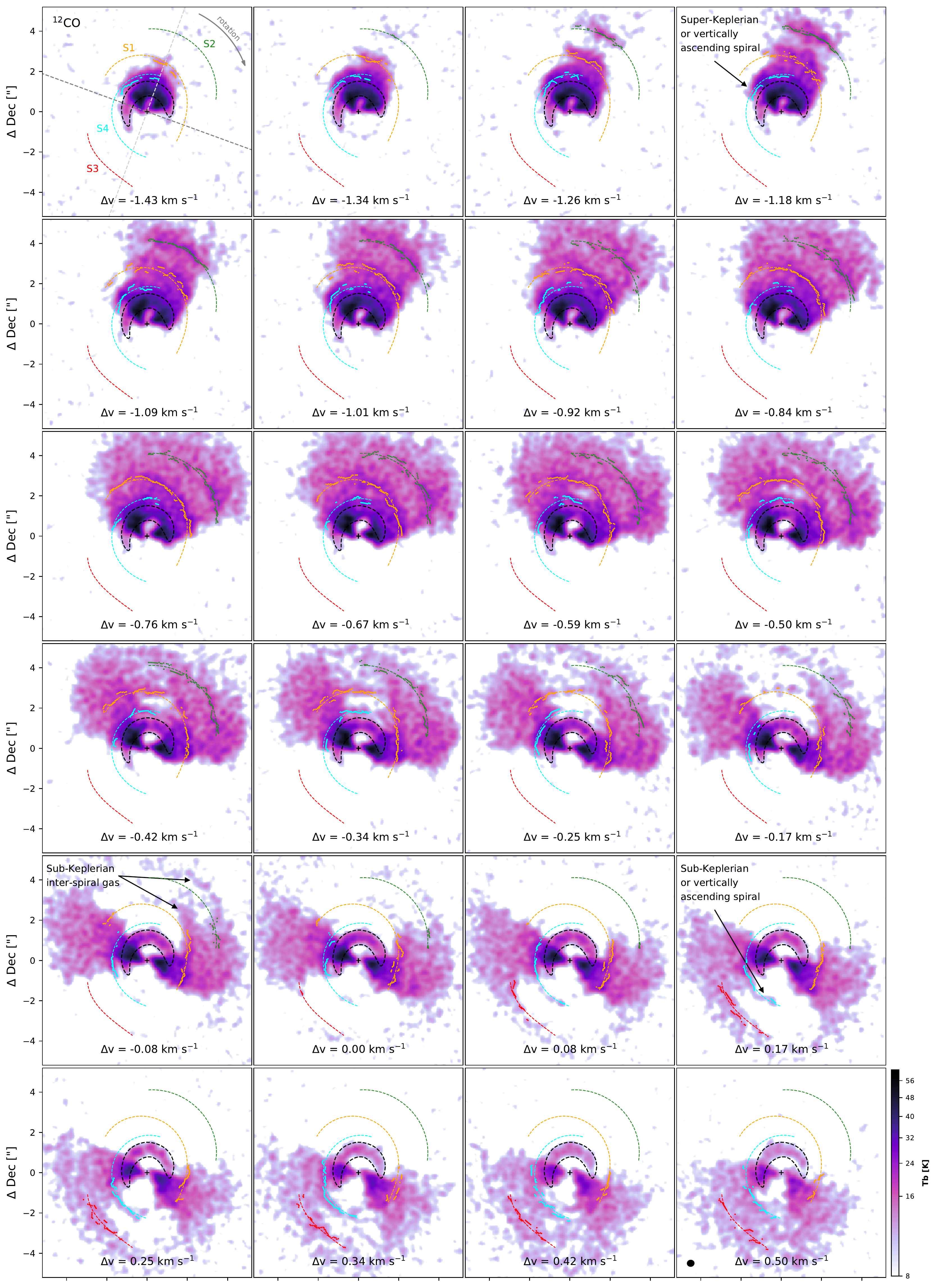}
    \caption{\textbf{Super-Keplerian spirals + sub-Keplerian inter-spiral gas in the outer disc.} Selected channel maps of $^{12}$CO without continuum subtraction. Each channel is 84\,m.s$^{-1}$ wide. The systemic velocity is 3.70\,km.s$^{-1}$. The \textit{light} and \textit{dark grey} dashed lines in the first panel represent the semi- major and minor axis respectively. The outer disc is inclined along PA = -20$^{\circ}$ such that NE is the far side whilst SW is the near side. The \textit{black} dashed contour represents the 13.6 mJy beam$^{-1}$ (0.2 $\times$ maximum) continuum emission. The colored dots represent components of the traced spirals. The colored dashed lines are the polynomial fits, of the form $r(\theta)=\sum_{i=0}^{3} a_{i}\theta^{i}$, determined using a least-squares fit on the spiral traces. The polynomial parameters are given in Appendix~\ref{sec:appendix2}. Beam is given on the bottom-right panels.}
    \label{fig:channels_12co}
\end{figure*}
\begin{figure*}
    %\ContinuedFloat
    %\centering
    \includegraphics[width=0.87\textwidth]{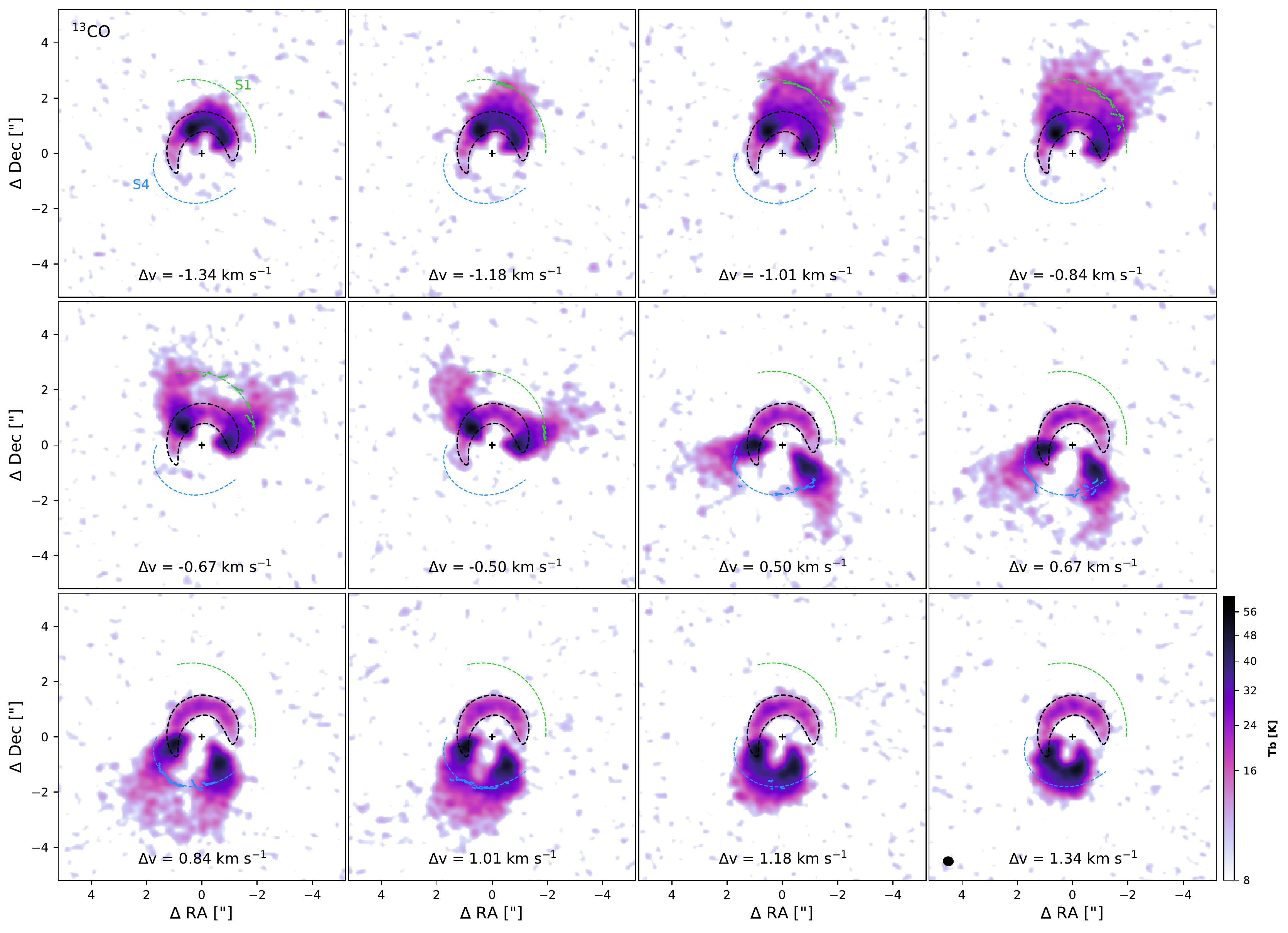}
    \caption{Selection of $^{13}$CO channel maps. The annotations are the same as in Fig.~\ref{fig:channels_12co}}
    \label{fig:channels_13co}
\end{figure*}

Figure \ref{fig:momentmaps} displays the different moment maps calculated from our observations. A warp in the central gas cavity is connecting to the inner disc in our moment 1\,(b), moment 2\,(c) and peak velocity\,(e) maps; confirming the twisted gas kinematics first reported by \cite{Casassus2015}. We also observe this feature in Figure~\ref{fig:twist}, where we show high velocity channels of the $^{12}$CO cube imaged using uniform weighting (resulting in a beam size of 0\farcs23 $\times$ 0\farcs25), to highlight the velocity field near the inner disc. A change in the position angle of the rotation axis for gas at high velocity is detected (PA\,$\approx$\,-55$^{\circ}$) in comparison to the average rotation axis of the outer disc (PA\,$\approx$\,70$^{\circ}$), \emph{i.e.} at extreme velocities, the rotation axis of the gas is offset by $\approx$\,125$^{\circ}$ clockwise. The measured velocities are consistent with Keplerian rotation around a 2.36\,$\pm$\,0.3\,\msun \citep{Boehler2021} star, but could also be tracing infalling material onto the inner disc \citep{Casassus2013, Rosenfeld2014}.

\subsection{Spiral arms in 12CO and 13CO}

\begin{figure*}
    %\centering
    \includegraphics[width=0.8\textwidth]{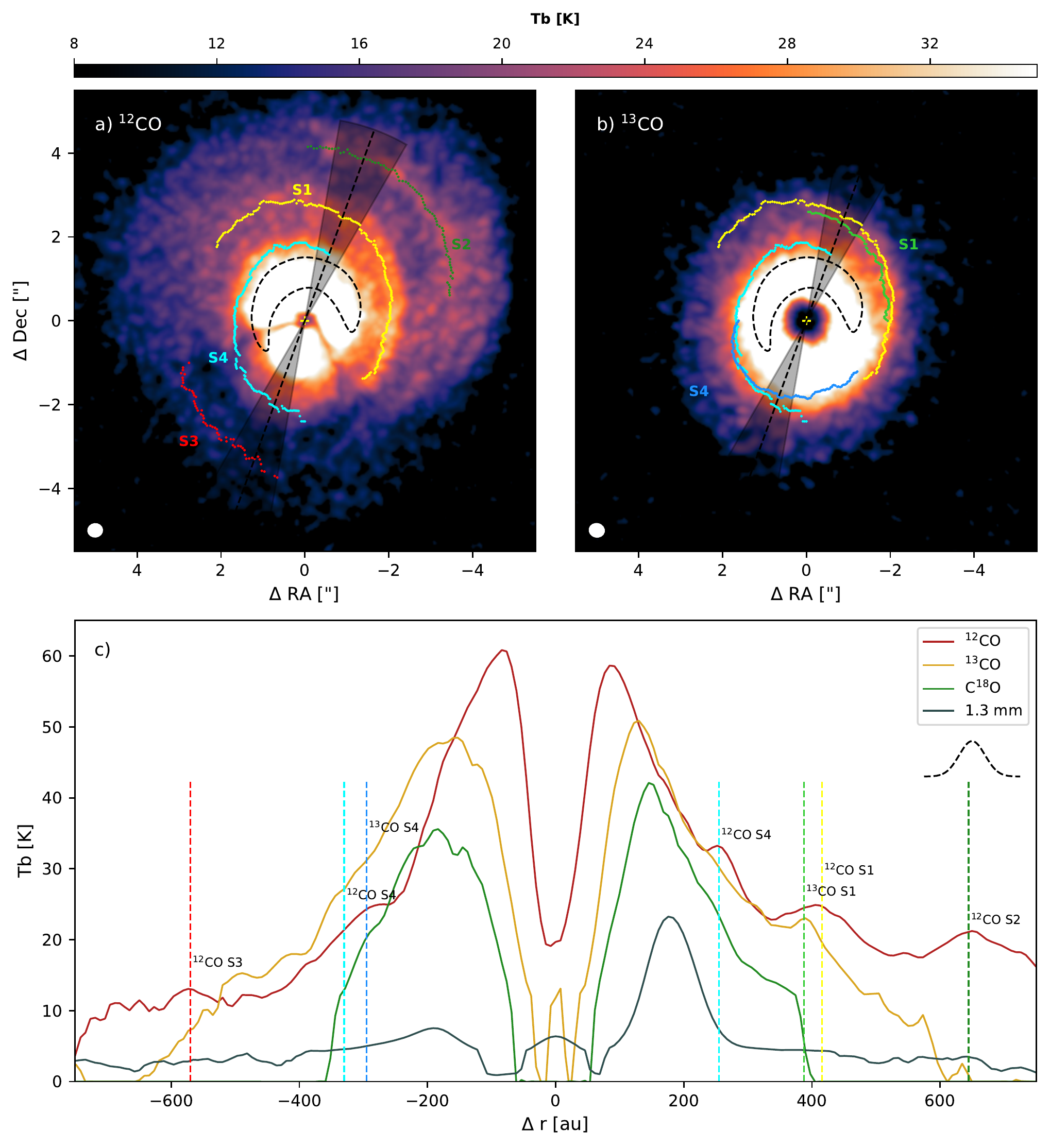}
    \caption{\textbf{Surfing spirals.} \textbf{a)} - \textbf{b)} Peak intensity maps for $^{12}$CO and $^{13}$CO, respectively, overlaid with spiral traces. The spirals labelled S1, S2 and S3 in $^{12}$CO are in agreement with their first reported detection by \citealp{Christiaens2014}. Counterparts to the aforementioned spirals are detected in $^{13}$CO, along with the new spiral: S4. Central emission has been saturated to highlight the the spiral structures, \emph{i.e.} a maximum T$_\mathrm{b}$ value of 35K has been applied on the colour bar. The transparent \textit{black} wedge between PAs -10$^{\circ}$ to -30$^{\circ}$ is used to compute a radial brightness temperature profile, with the diagonal line representing the outer disc's semi-major axis (PA=-20$^{\circ}$). The dashed \textit{black} contour represents the 13.6 mJy beam$^{-1}$ (0.2 $\times$ maximum) continuum emission. \textbf{c)} Radial brightness temperature profiles measured in the wedges. The vertical dashed lines represent the location of the spirals (with corresponding colours) as shown in panels \textbf{a} and \textbf{b}. The \textit{black}-dashed Gaussian represents the beam along the disc's semi-major axis.}
    \label{fig:spiraltrace+cut}
\end{figure*}

Figures \ref{fig:original}, \ref{fig:momentmaps}, \ref{fig:channels_12co} \ref{fig:channels_13co} reveal up to four large scale spiral structures in both $^{12}$CO and $^{13}$CO past the dust crescent. We trace these spirals in the peak intensity maps (Fig.~\ref{fig:spiraltrace+cut}) and channel maps by finding local maxima (>T$_\mathrm{b}$=8K for channel map traces and >0.3$\times$ maximum brightness temperature for intensity map traces of $^{12}$CO and $^{13}$CO) along radial cuts performed at 1$^{\circ}$ increments, for r > 210 au. The spirals labelled as S1, S2 and S3 in $^{12}$CO (e.g. Figs.~\ref{fig:channels_12co}) and \ref{fig:spiraltrace+cut}), match their previous reported detection by \cite{Christiaens2014}. We detect a new spiral (labelled S4) in $^{12}$CO originating from the dust crescent and potentially connecting up to S1, thus spanning more than 360$^\circ$. Furthermore, S1 may connect to S3, but low SNR between PA range $\sim$50--100$^{\circ}$ prevents a definitive conclusion. Similarly, it remains unclear whether S3 connects to S2 due to loss of emission towards SE caused by the foreground cloud (Fig.~\ref{fig:channels_12co}, bottom panel of the $^{12}$CO maps). Counterparts to two of the $^{12}$CO spirals (S1 and S4) are also observed in $^{13}$CO (Fig.~\ref{fig:channels_13co}).

Channel maps (Fig.~\ref{fig:channels_12co} and \ref{fig:channels_13co}) reveal the spirals not only in intensity, but also in velocity; as also reflected in velocity dispersion (Fig.~\ref{fig:momentmaps}c). The spirals extend to regions in the maps that correspond to either higher or lower velocities than the expected Keplerian rotation. For instance, S4 (cyan) at $\Delta$v = -1.10\,km.s$^{-1}$ (Fig.~\ref{fig:channels_12co}) is seen to extend towards the outer disc's semi-minor axis, \emph{i.e.} we detect emission in a region of the map where the velocity would be smaller if the disc was in Keplerian rotation. Conversely, both S4 and S3 in channel maps $\Delta$v = 0.08--0.25\,km.s$^{-1}$ are seen to extend towards the outer disc's semi-major axis, \emph{i.e.} into regions which would be at higher velocity if Keplerian. From this, S4 appears to be super-Keplerian in the blue-shifted regions of the map and sub-Keplerian in the red-shifted regions. Similarly, S3 appears to be sub-Keplerian. This trend is less clear for the S1 and S2 spirals. A possible explanation for the variation from super- to sub-Keplerian for S4 could alternatively be attributed to the spirals vertically ascending, whereby the Doppler-shift is attenuated in the red-shifted regions of the map whilst being amplified in the blue-shifted regions.

In contrast, the inter-spiral gas appears to be sub-Keplerian or vertically descending, \emph{i.e.} at $\Delta$v = -0.08\,km/s, we detect emission arcs between the traced spirals towards the semi-major axis, indicating that the inter-spiral gas is rotating at velocities smaller than the expected Keplerian velocity. These deviations in velocity are also apparent in the moment maps (in particular Fig.~\ref{fig:momentmaps}e; despite the cloud extinction that skews all integrated maps).

\begin{figure}
    \centering
    \includegraphics[width=\linewidth]{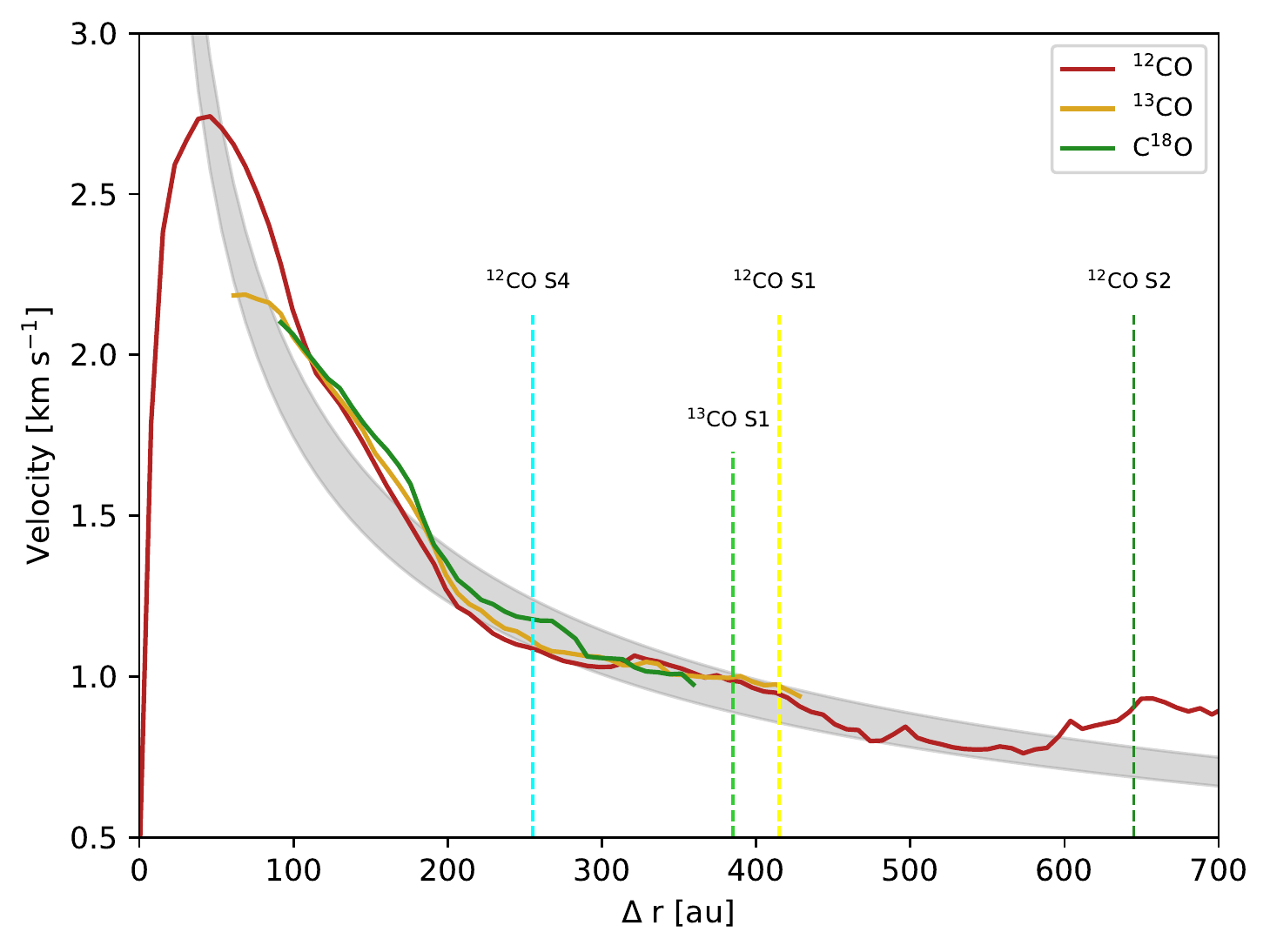}
    \caption{\textbf{Measured velocity profiles} for $^{12}$CO, $^{13}$CO and C$^{18}$O along the outer disc's semi-major axis, starting from disc centre and extending outwards in the NW direction. Systemic velocity of 3.70 km.s$^{-1}$ is subtracted from all velocity profiles. Velocities were extracted from the moment~1 maps. The \textit{light grey} curve represents Keplerian rotation at the disc midplane, where M$_\mathrm{star}$ = 2.36\,$\pm$\,0.3\,\msun\, and outer disc inclination~=~24$^{\circ}$.}
    \label{fig:velocity_profile}
\end{figure}

This is confirmed by a radial cut of the moment one map along the disc semi-major axis (Fig.~\ref{fig:velocity_profile}). The disc shows large non-Keplerian motions; oscillating between sub- to super-Keplerian motions. We compare the rotation curve of all three isotopologues ($^{12}$CO,$^{13}$CO and C$^{18}$O) to the expected midplane Keplerian rotation,
\begin{equation}
    v_{\mathrm{Kep}} = \sqrt{\frac{GM_{star}}{r}}\sin{(inc)}.
\end{equation}
with M$_\mathrm{star}$=2.36$\pm$0.3\msun \citep{Boehler2021}, G is the gravitational constant, r is the radial distance and \textit{inc}=24$^{\circ}$. Interestingly, we do not notice any significant variation of velocities between the isotopologues, as was for instance measured in IM Lupi \citep{Pinte2018}. Outside of 600\,au, the disc velocity deviates strongly from Keplerian rotation, suggesting the presence of a possible disc wind \citep[e.g.][]{Teague2019}. \cite{Huang2020} have also recently reported non-Keplerian motion in the enclosing gaseous disc of RU Lup.

In Figure~\ref{fig:momentmaps}f, we quadratically subtract the thermal velocity component ($\sqrt{2\,k\,T/m}$) from the local line width (integrated moment 2 map). For regions where $^{12}$CO is optically thick ($\tau$ > 3), the brightness temperature profile corresponds to the true gas kinematic temperature. However, in the outskirts of the disc, along with the central <35 au radial distance, where $^{12}$CO becomes optically thin the thermal velocity component is underestimated. The overall spiral structures seen in velocity dispersion, remain present post thermal component subtraction, and in combination with the non-Keplerian motion of both the spirals and the inter-spiral gas, imply that the observed spirals are not solely the result of local gas thermal motion but velocity perturbations likely caused by disc-binary interaction.

The shape of the spirals are subject to projection effects, which remain partially unknown due to uncertainty on the outer disc inclination (see Section~\ref{sec:intro}). To remedy this, we measure the distance to each spiral along the disc's semi-major axis, where projection effects are minimal, with a buffer of 10$^{\circ}$ on either side to minimise scatter \emph{i.e.} average distance in the wedge (Fig.~\ref{fig:spiraltrace+cut}a--b). The S4, S1 and S2 $^{12}$CO spirals are detected at a radial distance of $\sim$\,255, 415, and 645\,au in the NW direction, whilst towards the SE direction, the S4 and S3 $^{12}$CO spirals are located at $\sim$\,330 and 570 au, respectively.

The S1 $^{13}$CO spiral is detected at $\sim$\,385 au towards NW and the S4 $^{13}$CO spiral at $\sim$\,295 au towards SE. From these measurements, the $^{12}$CO and $^{13}$CO spirals are radially offset by approximately half-beam size ($\sim$30 au), with the $^{12}$CO spirals further out. In the NW quadrant this offset is approximately constant (Fig.~\ref{fig:spiraltrace+cut}b). We measure brightness temperatures of $\sim$25 and 23\,K for the $^{12}$CO and $^{13}$CO S1 spiral within the wedge. The spirals are detected outwards of the horseshoe where optically thick dust prevents stellar radiation from reaching the disc midplane, while the surface layers remain illuminated, as observed in scattered light \citep[e.g.][]{Avenhaus2014}. This will result in a vertical temperature gradient and thus a variation in the speed of propagation of the gas density wave, since $c_{s}\propto\sqrt{T}$. We hypothesize that the measured constant half-beam size radial offset between the two isotopologues may be due to the spirals \textit{surfing} along this vertical temperature gradient, whereby the spiral wave is propagating faster in the upper layers than at lower disc scale heights. \cite{Rosotti2020} detected a similar offset and variation of pitch angles between the spirals observed in scattered light and at sub-mm wavelengths in the disc surrounding HD~100453.

\subsection{Measuring the surface density inside the cavity}

\begin{figure*}
    \centering
    \includegraphics[width=\textwidth]{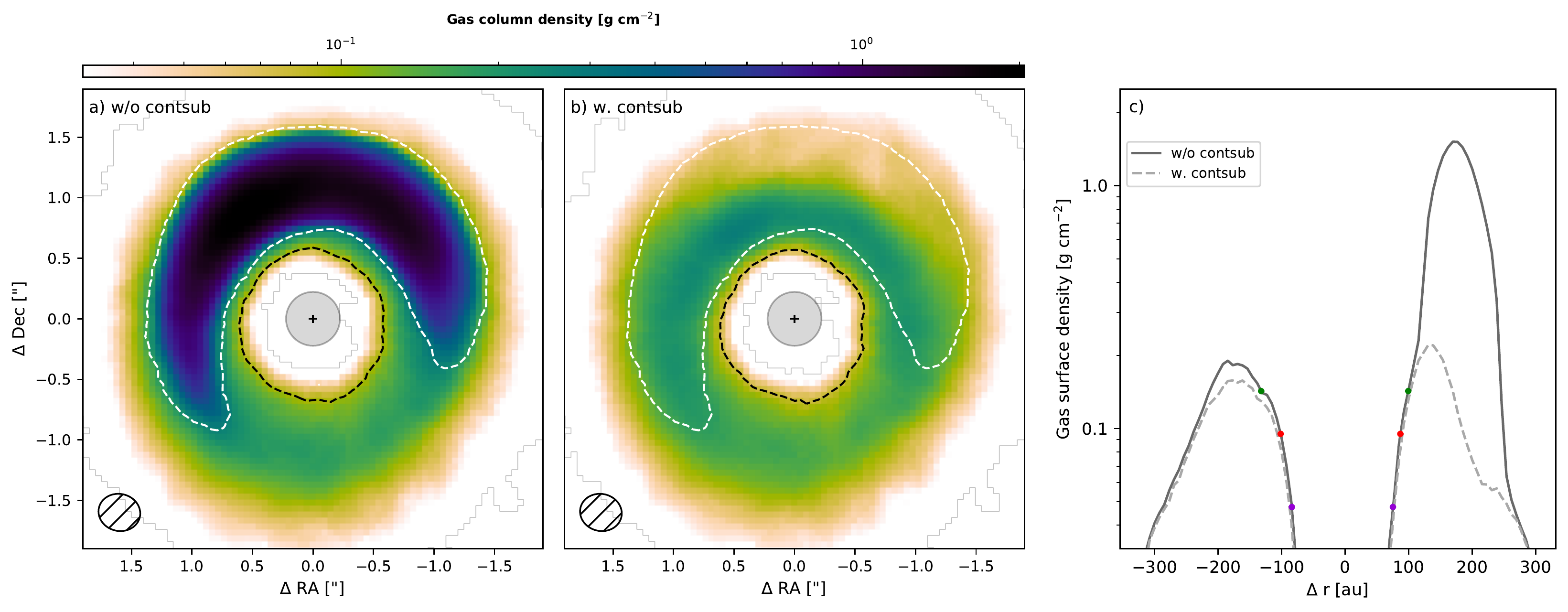}
    \caption{\textbf{Gas column density.} \textbf{a)}\,-\,\textit{b)} 2D gas column density maps without and with continuum subtraction, respectively. The \textit{black} dashed contour represents the inner half-maximum gas surface density. The \textit{white} dashed contours map the region where gas column densities computed with and without continuum subtraction differ by more than 50\%. The \textit{grey} circle represents the masked region, where all three isotopologues are optically thin. Beam is presented on the bottom-left corner. \textbf{c)} Gas surface density profiles with and without continuum subtraction along PA=-20$^{\circ}$ (outer disc's semi-major axis). A positive radial offset is directed towards NW and negative offset towards SE. The dots represent 1/2 max ($\sim$0.760 g\,cm$^{-2}$), 1/4 max ($\sim$0.380 g\,cm$^{-2}$) and 3/4 max ($\sim$1.140 g\,cm$^{-2}$) surface densities (red, purple and green, resp.) of the profile along the SE direction, without continuum subtraction.}
    \label{fig:surface_den}
\end{figure*}

A progressive increase in the radial extent of central gas depletion is observed from $^{12}$CO $\rightarrow$ $^{13}$CO $\rightarrow$ C$^{18}$O, as higher surface densities are required to reach the $\tau$=1 surface with decreasing isotopic abundance. We use this variation in optical depth to reconstruct a gas column density map from observations, where we require one optically thin tracer to compute column density, and one optically thick tracer to estimate local gas excitation temperature ($T_\mathrm{ex}$). Therefore, both tracers need to share the same $T_\mathrm{ex}$. Inside the scattered light cavity, near-infrared opacity is low, there is no vertical temperature gradient and the uniform slab approximation is valid. However, as dust becomes optically thick, this criteria is no longer satisfied. In this instance, we assume the $^{13}$CO and C$^{18}$O molecular surfaces are close in altitude (similar to the case with IM Lupi in \citealt{Pinte2018}) and thus share approximately the same excitation temperature.

The total molecular column density for an optically thin uniform slab is given by,
\begin{equation}
    N_{\rm mol} = N_{u} \frac{Z}{2J + 1}\exp{\left[ \frac{E_u}{kT_{\rm ex}}\right]},
\end{equation}
where $J=2$, $k$ is the Boltzmann constant, $E_u$ is the energy of the upper state and $Z$ is the partition function of a linear molecule. Assuming emission fills the beam, the column density in the upper level, $N_{u}$, is given by,
\begin{equation}
  N_{u} = \frac{4\pi}{h\nu A_{ul}}  \frac{B_\nu(T_\mathrm{ex})}{B_\nu(T_\mathrm{ex}) - B_\nu(T_\mathrm{bg})} \int (I_\nu -  I_{\nu, \mathrm{bg}})\, \mathrm{d}v,
\end{equation}
(equations 23 and 32 in \citealp{Mangum2015}) where $A_{ul}$ is the Einstein coefficient for spontaneous emission, $h$ is Planck's constant, $\nu$ is the rest frequency, $T_\mathrm{ex}$ is the excitation temperature and $I_\nu - I_{\nu, \mathrm{bg}}$ is the specific intensity minus the background, e.g. the quantity directly measured by the interferometer due to spatial filtering. Quantities for $E_u$ ($^{13}$CO = 15.87 K; C$^{18}$O = 15.81 K) and $A_{ul}$ ($^{13}$CO = 6.038 $\times$ 10$^{-7}$s$^{-1}$; C$^{18}$O = 6.011 $\times$ 10$^{-7}$s$^{-1}$) are taken from the Leiden Atomic and Molecular Database (LAMDA) \citep{Schoier2005} (accessed in July 2020).

The optical depth ($\tau$) for each isotopologue can be computed from the peak brightness temperature line ratio $R$ \citep[e.g.][]{Lyo2011},
\begin{equation}
    R = \frac{T_{B}(\nu_{1})}{T_{B}(\nu_{2})} = \frac{1 - e^{-\tau_{\nu_{1}}}}{1 - e^{-\tau_{\nu_{2}}}} = \frac{1 - e^{-\tau_{\nu_{1}}}}{1 - e^{-\tau_{\nu_{1}}/X}},
\end{equation}
where $\nu_{i=1,2}$ are the rest frequencies of either $^{12}$CO and $^{13}$CO or $^{13}$CO and C$^{18}$O pairs. X is the isotopic abundance ratio, which we assume to be ISM: [$^{12}$C]/[$^{13}$C] $\approx$ 70 \citep{Stahl2008} and [$^{16}$O]/[$^{18}$O] $\approx$ 500 \citep{Wilson1994}. We compute the various optical depths at each pixel in the map and find all three isotopologues to be optically thin < 35 au from disc centre, therefore we mask this region. In regions where the 3 isotopologues are detected (and not contaminated by the cloud), we find a similar $^{13}$CO optical depth when using the pairs $^{12}$CO/$^{13}$CO and $^{13}$CO/C$^{18}$O. Due to foreground cloud contamination of $^{12}$CO, we use $^{13}$CO to determine $T_\mathrm{ex}$ when $\tau_{^{13}\mathrm{CO}} > 3$, but $^{12}$CO otherwise. Similarly, $^{13}$CO is used to estimate column density when $\tau_{^{13}\mathrm{CO}} < 0.1$, but C$^{18}$O otherwise, as it remains optically thin throughout the disc. Gas emission within the horseshoe suffers from dust contamination, and vice-versa; continuum emission close to line centre is absorbed by the gas and performing continuum subtraction largely underestimates line flux due to overestimating dust emission contribution \citep{Boehler2017}. Therefore, we use integrated flux ($I_{\mathrm{int}}$) maps with and without continuum subtraction to place a lower and upper limit, respectively, on the gas column density inside the horseshoe (refer to Fig.~\ref{fig:surface_den}). The gas column densities are scaled by assuming a constant ISM abundance ratio of [$^{12}$CO]/[H$_{2}$] $\approx$ 10$^{-4}$, and the isotopic ratios above. This is assuming no isotopologue-selective photodissociation, freeze-out or chemical evolution is altering the CO isotopologue's abundance levels, the implications of which are discussed in section~\ref{sec:origin}.

Towards the SE direction along the disc's semi-major axis (refer to Fig.~\ref{fig:surface_den}c), where emission is not heavily contaminated by dust grains, we measure a gas cavity size at half-maximum surface density of $\sim$100 au ($\sim$0.760 g\,cm$^{-2}$). We find a steep surface density profile at the edge of the cavity with the one-quarter and three-quarter maximum gas surface densities reached at $\sim$85 ($\sim$0.380 g\,cm$^{-2}$) and $\sim$130 ($\sim$1.140 g\,cm$^{-2}$) au, respectively. Our measured gas cavity size is $\sim$3 to 8 times larger than the semi-major axis of the companion ranging between 12 to 31 au \citep{Claudi2019}. Deprojecting the generated column density map at estimated disc inclinations of 20, 24 and 28$^{\circ}$, we measure an eccentricity of 0.32 (PA$\sim$\,-43$^{\circ}$), 0.35 (PA$\sim$\,-53$^{\circ}$) and 0.44 (PA$\sim$\,-59$^{\circ}$) respectively for the gas cavity by fitting an Ellipse to the half-maximum surface density contour (black dashed contour) shown in Figure~\ref{fig:surface_den}. An eccentric gas cavity is suggestive of dynamical clearing by the binary.

Hydrodynamical simulations predict cavity sizes of 2-3 times the binary semi-major axis for co-planar orbits \citep{Artymowicz1994}, up to a factor of 5 for integration times reaching a few viscous spreading times \citep[e.g.][]{Hirsh2020} and potentially up to 7 times \citep{Thun2017}. Highly inclined binary orbits respective to the disc, as in HD 142527, result in smaller cavities however, typically twice the binary semi-major axis \citep{Price2018, Hirsh2020}. Our measured cavity size suggests potential tension with current models of binary-disc interaction, although this can only be confirmed with more precise orbital constraints.

\subsection{Origin of the dust horseshoe}
\label{sec:origin}

Three hypotheses have been proposed to explain asymmetric dust emission: i) dust trapping in a vortex due to a planet on a circular orbit \citep{Birnstiel2013, Lyra2013}, ii) a traffic jam caused by eccentric orbital motion \citep{Ataiee2013}, and iii) dust trapping in an horseshoe/`over-dense lump' caused by a massive companion \citep{Farris2014, Ragusa2017, Ragusa2020, Calcino2019}.

Our reconstructed column density profiles demonstrate that regardless of continuum subtraction, gas column density peaks $\sim$2--10 times higher within the dust horseshoe than on the opposite side, consistent with the modelling results of \cite{Muto2015}. This is lower than the contrast in dust emission of $\sim$30, as measured from our continuum data, to $\sim$40 derived by modelling the continuum emission by \cite{Soon2017}. Here we assume a continuous dust distribution in the horseshoe, where sub-millimetre dust grains have been shown to be optically thick\citep{Casassus2015b}, therefore the 1.3\,mm emission is expected to be dominated by grains of $\approx200\,\mu$m radius (\emph{i.e.} $2\pi\,a\approx\lambda$). For our lower and upper estimates of gas surface density within the horseshoe (0.2 and 1.5\,g\,cm$^{-2}$, refer to Fig.~\ref{fig:surface_den}), the resulting Stokes number for the dust grains ranges between 0.2 to 2 \citep[e.g.][]{Dipierro2015a}, assuming spherical grains and a typical grain density of 3.5\,g\,cm$^{-3}$. This means that the dust grains dominating the emission in the crescent are the ones experiencing the most efficient gas drag. Several factors, such as isotope-selective photodissociation, freeze-out and/or chemical evolution in PPDs, can drive CO isotopologue abundance levels away from ISM values \citep{Miotello2014, Miotello2016}. Here, we have made no assumption on the extent of such depletion by the aforementioned processes, as the exact level of carbon/oxygen depletion in the disc is unknown. The corresponding Stokes number would scale inversely with the depletion factor.

Based on our upper and lower limits on the Stokes number, the horseshoe is likely the result of radial and azimuthal trapping of dust grains, ruling out the traffic jam scenario which predicts no dust trapping \citep{Ataiee2013}. Our measurement of an eccentric gas cavity rules out the scenario of a planet on a circular orbit in the outer disc. The only remaining possibility is that a dust-trapping vortex might be created by an additional planetary-mass companion on a stable eccentric orbit. This leaves model iii) as the most likely scenario, namely dust trapping in a gas over-density created by the binary (as already demonstrated by \citealt{Price2018}).

\section{Conclusion}

Our ALMA band 6 observations of HD~142527 show:
\begin{enumerate}
    \item four spiral arms extending as far as 650~au in $^{12}$CO and two counterparts to these spirals in $^{13}$CO up to 400~au. Three of the four $^{12}$CO spirals had been previously reported by \cite{Christiaens2014}, here we report a new $^{12}$CO spiral originating from the horseshoe along with detections in $^{13}$CO. These spirals are non-Keplerian in nature (either super/sub-Keplerian, vertically ascending or a combination of both) and display a locally larger linewidth (Fig.~\ref{fig:momentmaps}c);
    \item a radial offset of approximately half-beam size ($\sim$30 au) for the S1 and S4 spiral arms traced between $^{12}$CO and $^{13}$CO (where $^{12}$CO is further out); similar to the offsets seen in HD~100453 by \cite{Rosotti2020}. We hypothesize this spatial offset as the result of a variation in propagation, at sound speed, due to a vertical temperature differential in the outer regions;
    \item spatially resolved central gas depleted emission and varying optical depths in all three gas tracers, allowing us to reconstruct a 2D gas column density map. From this, a gas cavity size at half-maximum surface density of $\sim$100\,au is measured with a steep radial profile. The shape of the gas cavity edge has eccentricity ranging from 0.3 to 0.45;
    \item a gas density concentration in the horseshoe, where we derive a Stokes number of $\sim$1 for sub-mm sized dust grains. The contrast in gas and dust density distribution is in agreement with previous modelling efforts by \cite{Muto2015} and \cite{Soon2017}, respectively. This supports the hypothesis of dust trapping being the origin of the unique horseshoe morphology \citep{Ragusa2017,Ragusa2020} and with the most likely origin being a gas pressure maximum created by the binary companion \citep{Price2018}.
\end{enumerate}

The current spatial resolution prevents reaching a definitive conclusion on the connection between the various spirals. Observations at 0\farcs1 with $\approx$ 8h of ALMA time would achieve a similar signal-to-noise ratio, but allow to pinpoint the complex structure of the spiral pattern in the disc.

%###########################################################################

\section*{Acknowledgements}
This paper makes use of the following ALMA data:
 and ADS/JAO.\-ALMA\#2015.1.01353. ALMA is a partnership of ESO (representing
 its member states), NSF (USA) and NINS (Japan), together with NRC (Canada),
 MOST and ASIAA (Taiwan), and KASI (Republic of Korea), in cooperation with the
 Republic of Chile. The Joint ALMA Observatory is operated by ESO, AUI/NRAO and
 NAOJ. The National Radio Astronomy Observatory is a
facility of the National Science Foundation operated under cooperative
agreement by Associated Universities, Inc. C.P., D.J.P. and V. C. acknowledge funding from the Australian Research Council via FT170100040, FT130100034, and DP180104235. S.M. is supported by a Reasearch Fellowship from Jesus College, University of Cambridge. Y.B. acknowledges
funding from ANR (Agence Nationale de la Recherche) of France under contract number ANR-16-CE31-0013 (Planet-Forming-Disks). S.C. and S.P. acknowledge FONDECYT grants 1171642 and 1191934.

\newpage
\section*{Data availability}
Reduced and calibrated data cubes are available at https://doi.org /10.6084/m9.figshare.13625144.v1. The original raw data is publicly available via the ALMA archive under Project code: 2015.1.01353.S.

%%%%%%%%%%%%%%%%%%%%%%%%%%%%%%%%%%%%%%%%%%%%%%%%%%



\bibliographystyle{mnras}
\bibliography{ref} % if your bibtex file is called example.bib

%\begin{thebibliography}{99}
%\end{thebibliography}

%%%%%%%%%%%%%%%%%%%%%%%%%%%%%%%%%%%%%%%%%%%%%%%%%%

%%%%%%%%%%%%%%%%% APPENDICES %%%%%%%%%%%%%%%%%%%%%

\appendix

\section{cloud contamination and correction}
\label{sec:appendix}

$^{12}$CO emission in the Southern region is partially filtered out by the interferometer due to the presence of the molecular cloud at velocities between 3.0 to 5.7\,km/s \citep{Casassus2013b}. As a result, signal is attenuated at the aforementioned velocities. This however does not affect the shape of the emission in individual channels, but only the flux. A caveat of this spatial filtering is the skewing of moment maps, such as in Figures ~\ref{fig:original} and \ref{fig:momentmaps}, which complicates their interpretation.

\begin{figure*}
    \centering
    \includegraphics[width=\textwidth]{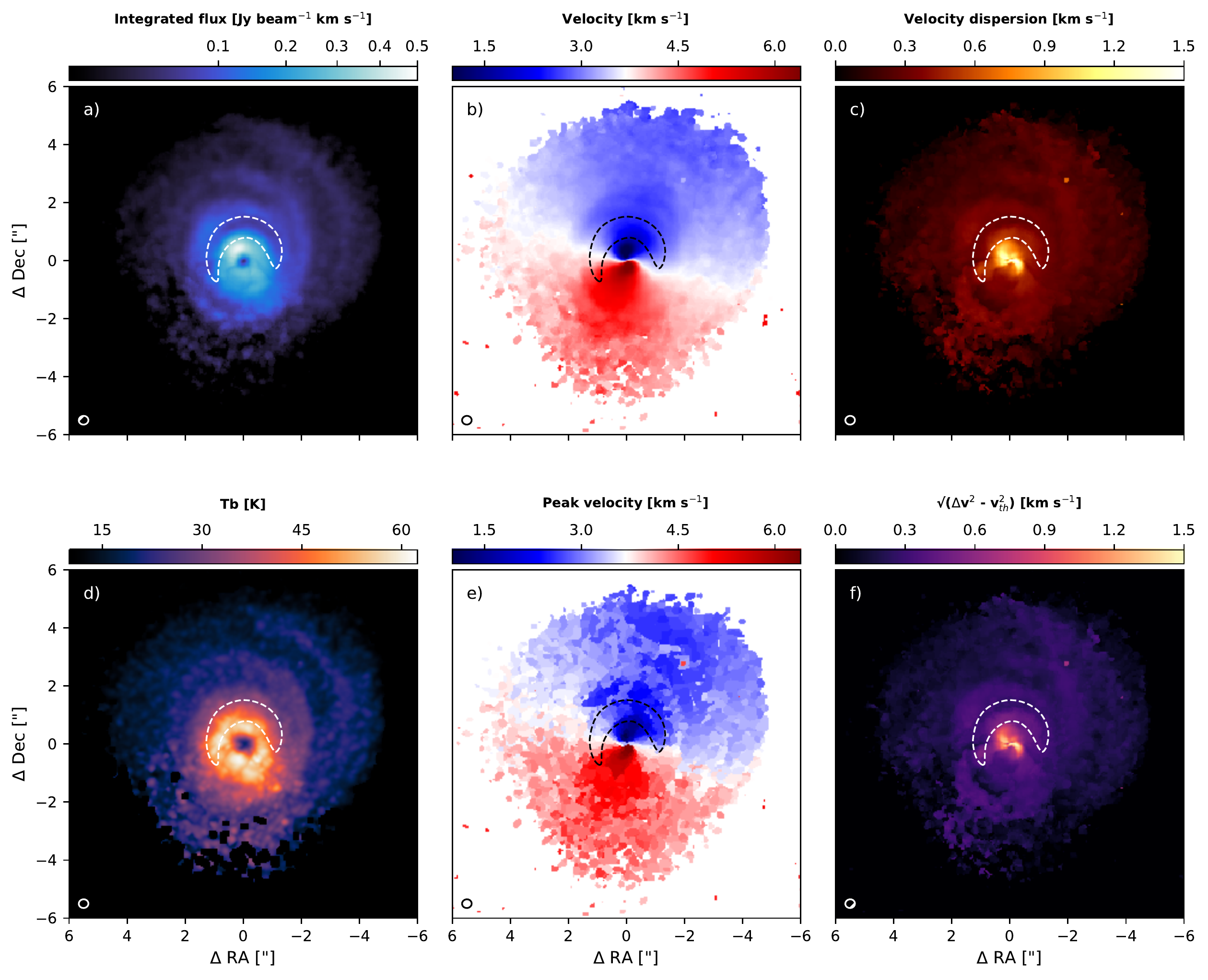}
    \caption{\textbf{Cloud corrected moment maps.} \textbf{a)}--\textbf{f)}, moment 0, moment 1, moment 2, peak brightness temperature, peak velocity and velocity dispersion minus thermal component for $^{12}$CO. Emission between 3.0--5.7\,km/s have been corrected for extinction from the intervening cloud. All integrated maps, with the exception of peak brightness temperature, are continuum subtracted. The dashed contours represents the 13.6 mJy beam$^{-1}$ (0.2 $\times$ maximum) continuum emission. Beam is given on the bottom left.}
    \label{fig:cloudcorrected}
\end{figure*}

We attempt to correct the extinct channels, by matching the $^{12}$CO J=2-1 line profile to the J=6-5 profile observed by \cite[][Fig 3]{Casassus2015}. We correct the J=2-1 with a Gaussian profile centred at $v=4.54$km/s, with a FWHM of $1.1$km/s and an optical depth of 2.3 at line center. In turn, this confirms that the cloud contamination for the $^{13}$CO and C$^{18}$O isotopologues is negligible. This correction is of course highly uncertain, but allows us to produce moment maps that highlight, at a higher signal-to-noise, the spiral structures we detect in individual channel maps towards the SE region (Fig.~\ref{fig:cloudcorrected}).

\section{best-fit parameters to spiral arm traces}
\label{sec:appendix2}

\begin{table}
    \caption{\textbf{Spiral arm fit parameters.} Coefficients for the polynomial, of the form $r(\theta)=\sum_{i=0}^{3} a_{i}\theta^{i}$, fit to the spiral structures (S1, S2, S3 and S4) traced in $^{12}$CO and $^{13}$CO using a least-squares fit.}
    \begin{tabular}{lcccr}
    \hline
    \hline
    $^{12}$CO & & & &\\
    \hline
    \vspace{2mm}
    & S1 & S2 & S3 & S4\\
    a$_\mathrm{0}$ & 546 & -134 & -277 & 37.9\\
    a$_\mathrm{1}$ & -294 & 36.8 & 435 & -2.31\\
    a$_\mathrm{2}$ & 55.2 & 5.95 & -185 & -0.42\\
    a$_\mathrm{3}$ & -3.31 & -0.99 & 26.5 & 0.64\\
    \hline
    \hline
    $^{13}$CO & & & &\\
    \hline
    \vspace{2mm}
    & S1 & S2 & S3 & S4\\
    a$_\mathrm{0}$ & 496 & -- & -- & -50.5\\
    a$_\mathrm{1}$ & -258 & -- & -- & 97.6\\
    a$_\mathrm{2}$ & 47.2 & -- & -- & -34.1\\
    a$_\mathrm{3}$ & -2.75 & -- & -- & 3.78\\
    \hline
    \end{tabular}
    \label{tab:spiralfits}
\end{table}

%%%%%%%%%%%%%%%%%%%%%%%%%%%%%%%%%%%%%%%%%%%%%%%%%%

% Don't change these lines
\bsp	% typesetting comment
\label{lastpage}
\end{document}